\documentclass[sigconf]{acmart}

\AtBeginDocument{%
  \providecommand\BibTeX{{%
    \normalfont B\kern-0.5em{\scshape i\kern-0.25em b}\kern-0.8em\TeX}}}

\setcopyright{acmcopyright}
\copyrightyear{2023}
\acmYear{2023}
\acmDOI{XXXXXXX.XXXXXXX}

\acmPrice{15.00}
\acmISBN{978-1-4503-XXXX-X/18/06}

\usepackage{graphicx,subcaption,ragged2e}
\usepackage{tabularx} 
\usepackage{xcolor}
\usepackage{multirow}
\usepackage{multicol}
\usepackage{lineno}

\newcommand{\tableref}[1]{Table~\ref{table:#1}}

\newcommand{\figref}[1]{Figure~\ref{fig:#1}}

\definecolor{pblue}{rgb}{0.13,0.13,1}
\definecolor{pgreen}{rgb}{0,0.5,0}
\definecolor{pred}{rgb}{0.9,0,0}
\definecolor{pgrey}{rgb}{0.46,0.45,0.48}
\definecolor{gray}{rgb}{0.4,0.4,0.4}
\definecolor{darkblue}{rgb}{0.0,0.0,0.6}
\definecolor{cyan}{rgb}{0.0,0.6,0.6}

\begin{document}


\title[Simpson's Paradox and Completion Trends]{
Simpson's Paradox and Lagging Progress in Completion Trends of Underrepresented Students in Computer Science}







\author{John Mason Taylor}
\email{john.taylor@furman.edu}
\affiliation{%
  \institution{Furman University}
  \city{Greenville}
  \state{SC}
  \country{USA}
  \postcode{29613}
}

\author{Rebecca Drucker}
\email{rebecca.drucker@furman.edu}
\affiliation{%
  \institution{Furman University}
  \city{Greenville}
  \state{SC}
  \country{USA}
  \postcode{29613}
}

\author{Chris Alvin}
\email{chris.alvin@furman.edu}
\affiliation{%
  \institution{Furman University}
  \city{Greenville}
  \state{SC}
  \country{USA}
  \postcode{29613}
}

\author{Syed Fahad Sultan}
\email{fahad.sultan@furman.edu}
\affiliation{%
  \institution{Furman University}
  \city{Greenville}
  \state{SC}
  \country{USA}
  \postcode{29613}
}


\begin{abstract}
It is imperative for the Computer Science (CS) community to ensure active participation and success of students from diverse backgrounds. This work compares CS to other areas of study with respect to success of students from three underrepresented groups: Women, Black and Hispanic or Latino. Using a data-driven approach, we show that trends of success over the years for underrepresented groups in CS are lagging behind other disciplines. Completion of CS programs by Black students in particular shows an alarming regression in the years 2011 through 2019. This national level decline is most concentrated in the Southeast of the United States and seems to be driven mostly by a small number of institutes that produce a large number of graduates. We strongly believe that more data-driven studies in this area are necessary to make progress towards a more equitable and inclusive CS community. Without an understanding of underlying dynamics, policy makers and practitioners will be unable to make informed decisions about how and where to allocate resources to address the problem.
\end{abstract}




\maketitle

\section{Introduction}
\label{sec:introduction}

The Computer Science (CS) student body of today is going to build and shape our futures for tomorrow. To ensure that the benefits of future technology extend to all, it is important that the field of CS is inclusive, equitable and reflects the diverse perspectives and talents of all individuals.

Diversity, Equity, and Inclusion (DEI) in CS is a topic of growing interest in the United States~\cite{denner2023equity, sax2017anatomy, ibe2018reflections}. 
However, in our review of the literature, we found a dearth of DEI studies in CS that take a data-driven approach. Most of the existing work is qualitative and prescriptive in nature~\cite{kulkarni2018promoting, larsen2005increasing}. 
To address this gap in Computing Education literature, in this work, we use 9 years of Integrated Post-secondary Education Data System (IPEDS) Completions \cite{statistics2012integrated} data to better understand key trends in success or struggles of underrepresented students in Computer Science. 


Our results show that the completion trends in CS are lagging behind those in other areas of study (i.e., non-CS) for students from three key underrepresented demographic groups: Women, Black and Hispanic or Latino. For Black students, we observe an alarmingly sharp decline in CS completions during the same period when no such decline is observed in other areas of study. We show that this is despite the fact that most institutions are individually exhibiting a positive trend in CS completions for Black students.
This is an example of Simpson's Paradox: a phenomenon of different or opposite patterns observed at two different scales.
We explain that a few institutions with a high number of total completions disproportionately influence national level trends in CS completions for Black students.
Negative trends in CS for Black students are observed to be concentrated primarily in the Southern United States with a large cluster in the Southeast region. 

For Women, we observe that the completions in CS are far behind completions in other areas of study but the trend over the years is consistently positive. For Hispanic or Latino students, we report that even though completion trends are positive, they are still lagging behind trends observed in other areas of study.
Altogether, our results strongly suggest a call to action for more data-driven research of underrepresented groups in CS and the broader topic of DEI in CS. Only with a better understanding of the underlying dynamics can we hope to make progress toward a more equitable and inclusive CS community.

%
%
\vspace{-0.1in}
\section{Methodology}

\textbf{Data.}
The Integrated Postsecondary Education Data System (IPEDS) is a set of interrelated surveys conducted annually by the National Center for Education Statistics (NCES), a part of the Institute for Education Sciences within the United States Department of Education.
The \emph{Completions survey} within this system is designed to collect information on the number and types of degrees awarded by all accredited U.S. postsecondary institutions as well as information on the characteristics of degree recipients.
The data is publicly available to download from the IPEDS website \cite{statistics2012integrated}. 

In this work we analyze $9$ years of data (2011-2019) from the Completions survey to understand the success of underrepresented students in CS.
These nine years were selected to minimize any confounding effects that might be attributed to the financial crisis starting in 2008 and the COVID-19 pandemic beginning in 2020.

The data we use in this paper are from the survey table titled \emph{Awards/degrees conferred by program, award level, race/ethnicity, and gender}.
We specifically focus on undergraduate level degrees (Award Level code $5$).
We define `Computer Science' as degrees awarded under Classification of Instructional Programs (CIP) code $11$ titled \emph{Computer and Information Sciences and Support Services} in the data.

\noindent \textbf{Analyses.}
Our analyses focus on the following three underrepresented groups: Women (\texttt{variable=CTOTALW}), Black (\texttt{variable=} \texttt{CBKAAT}) and Hispanic or Latino (\texttt{variable=CHISPT}).
Values for these three variables are self-reported by students and communicated to NCES by institutions and are in the IPEDS Completions survey tables at the institutional level.
With the gender category, there was no way to report anything other than ‘Men’ and ‘Women’ in IPEDS prior to $2022$.
A question was added in the 2022-23 survey to capture the total number of non-binary students and the total number of students for whom gender is unknown \cite{ipeds-faq}.
We acknowledge the limitations of this binary definition of gender for past data and welcome expansion of gender categories going forward.

%
%
Counts were computed using a sum of the relevant variable over all institutions for a given year. Percentages were computed by dividing relevant sums by \texttt{CTOTALT}: total of students awarded degrees with \texttt{CIPCODE} in range $[11, 12)$.

We use the Pearson correlation coefficient $r$ between \emph{count of completions} with years $[2011, 2019]$ to measure trends over time; similarly for \emph{percentage of completions} over years.
$r$ measures how two continuous variables co-vary and indicate a linear relationship as a number between $-1$ (perfectly negatively correlated) to $0$ (not correlated) to $1$ (perfectly positively correlated).







We use Kernel Density Estimation (KDE) to identify regions of high density in bivariate visualizations.
Given a random variable $v$, KDE estimates the probability density function (PDF) non-parametrically.
Conceptually, KDE is a smoothed version of a histogram for $v$.

\noindent \textbf{Source code.}
Complete analyses and source code is available at \url{https://github.com/REDACTED}.

%
%

\begin{figure}
\captionsetup[subfigure]{justification=Centering}
\begin{subfigure}[t]{\columnwidth}
    \centering
    \includegraphics[width=\textwidth]{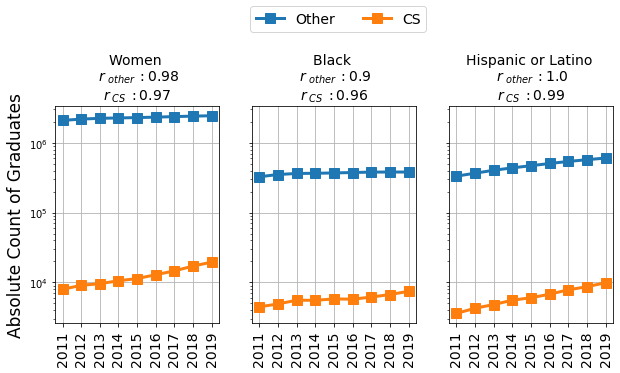}
    \caption{Absolute count of completions. These data have been log-scaled and all vertical axes use the same scale.}
    \label{fig:national_a}
\end{subfigure}%
\hspace{\fill} 
\begin{subfigure}[t]{\columnwidth}
    \centering
    \includegraphics[width=\columnwidth]{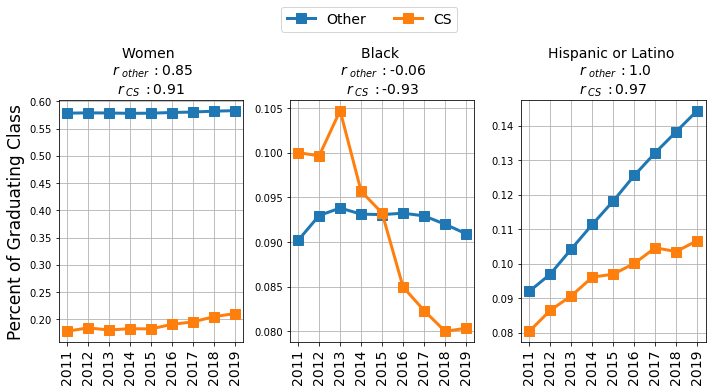}
    \caption{Percentage of total completions}
    \label{fig:national_b}
\end{subfigure}
\caption{National completion trends in CS and other areas of study from 2011 to 2019 for Women, Black, and Hispanic or Latino students.}
\label{fig:national}
\end{figure}

\begin{figure*}[t!]
\captionsetup[subfigure]{justification=Centering}
\begin{subfigure}[t]{.3\textwidth}
    \includegraphics[width=\textwidth]{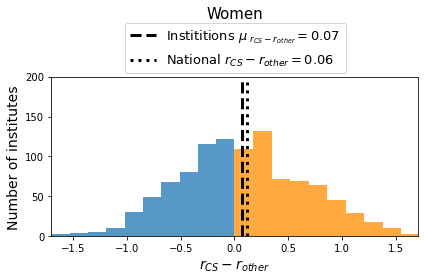}
\end{subfigure}%
\hspace{\fill}
\begin{subfigure}[t]{.3\textwidth}
    \includegraphics[width=\linewidth]{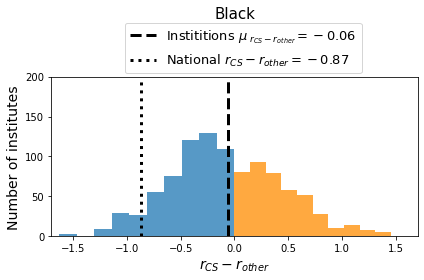}
\end{subfigure}%
\hspace{\fill}
\begin{subfigure}[t]{.3\textwidth}
    \includegraphics[width=\linewidth]{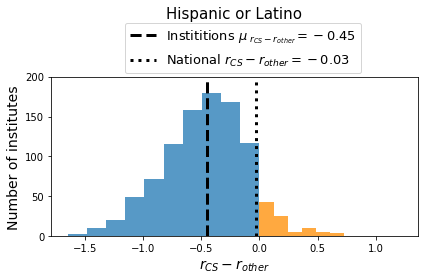}
\end{subfigure}
\vspace{-0.25cm}
\caption{Comparing CS completion trends with completion trends in other areas of study using $r_{CS} - r_{other}$.
Each dashed line is a mean for the corresponding distribution ($\mu_{r_{CS} - r_{other}}$) while each dotted line corresponds to National statistics of $r_{CS} - r_{other}$ as reported in \figref{national_b}.
Frequencies of all institutions with $r_{CS} - r_{other} > 0$ are colored orange; $r_{CS} - r_{other} < 0$ are blue.}
\label{fig:diffs}
\end{figure*}

\raggedbottom
\section{Results}
\label{seg:results}

In this section, we present our findings from analysis of 9 years of IPEDS completions data in Computer Science.
In subsection \ref{sec:abs_relative}, we argue for the use of relative counts (i.e. percentages) over absolute counts by demonstrating how completion trends using percentages are more informative and meaningful. 
In subsection \ref{sec:lagging_progress}, we demonstrate the lagging progress of Women, Black and Hispanic or Latino students in CS when compared to other areas of study (i.e., non-CS).

We continue in subsection \ref{sec:simpson} by comparing trends observed at the national level to trends observed at the level of individual institutions.
Our results demonstrate that for Black and Hispanic or Latino students, completion trends disappear or reverse at these two scales thus indicating the occurrence of Simpson's paradox.
We then attempt to explain underlying reasons behind the paradox. 

Finally, we present the geographic distribution of completion trends in CS across the United States, reporting spatial clusters of declining trends for students from underrepresented groups.

\subsection{Absolute vs. Relative Counts}
\label{sec:abs_relative}
\begin{table*}[]
\caption{National-level and Institutional-level completion trends for CS and other areas of study.
}
\vspace{-0.25cm}
\begin{tabular}{|c|c|c|c|cc|}
\hline
\multirow{2}{*}{\textbf{}} & \multirow{2}{*}{Demographic} & $r_{CS}$ & $r_{other}$ & \multicolumn{2}{c|}{$r_{CS} - r_{other}$} \\ \cline{3-6} 
 &  & National & National & \multicolumn{1}{c|}{National} & Institutional Average \\ \hline
1 & Women & 0.91 & 0.85 & \multicolumn{1}{c|}{0.06} & 0.07 \\ \hline
2 & Black & -0.93 & -0.06 & \multicolumn{1}{c|}{-0.87} & -0.06 \\ \hline
3 & Hispanic or Latino & 0.97 & 1 & \multicolumn{1}{c|}{-0.03} & -0.45 \\ \hline
\end{tabular}
\label{table:diffs}
\end{table*}

%
%

In comparing CS with other areas of study, we are interested not only in the gap between completions for students from underrepresented groups, but also in the trends over time.

\figref{national} depicts national level trends in CS completions as they compare to other areas of study (Non-CS).
\figref{national_a} presents completion trends computed as absolute counts while \figref{national_b} completion trends with percentages.
Absolute counts are computed as the sum of all completions for a given year and demographic group.
It can be observed that when computed as absolute counts, completions across all underrepresented groups exhibit strong positive trends for all demographic groups in CS as well as other areas of study.

Used as a measure of participation or success, absolute counts do not lead to fair comparisons.
In \figref{national_a}, CS completions are compared to total completions for \textit{all other} areas of study.
The wide gap between the \texttt{CS} line (orange) and the \texttt{other} line (blue) is by definition; hence, the vertical axes of all panels in~\figref{national_a} are log-scaled. 
\figref{national_b} presents completion trends as percentages, computed as the sum of completions by students from a demographic group in a given year divided by the sum of all completions that year.
It is now the case in \figref{national_b} that the gap between the trend lines for the two groups is meaningful.
It is also the case that the trend lines individually exhibit more interesting patterns in stark contrast to absolute counts in \figref{national_a}.

When measured using absolute counts, CS exhibits positive trends for all three underrepresented groups as shown in \figref{national_a}.
However, when using relative counts, the CS trends weaken, and in the case of Black students, a reversal occurs.
We conclude that even though CS is seeing an increase in the absolute count of completions overall, the \emph{share of underrepresented students in these completions is not increasing at the same rate}, if at all.
In fact, as shown in \tableref{diffs}, the share of Black students in CS completions experienced a sharp decline $r_{CS \wedge Black} = -0.93$ during the same time when their share in other areas of study remained relatively stable $r_{other \wedge Black} = -0.06$.

\begin{figure*}[t!]
\centering
\captionsetup[subfigure]{justification=Centering}
\begin{subfigure}[t]{.3\textwidth}
    \includegraphics[width=\textwidth]{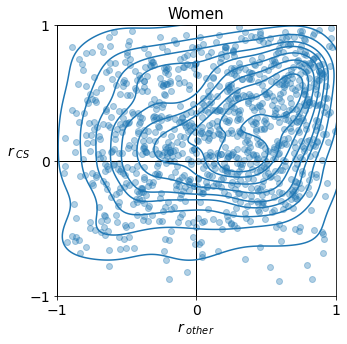}
\end{subfigure}%
\hspace{\fill}
\begin{subfigure}[t]{.3\textwidth}
    \includegraphics[width=\linewidth]{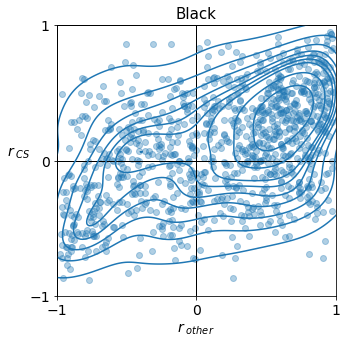}
\end{subfigure}%
\hspace{\fill}
\begin{subfigure}[t]{.3\textwidth}
    \includegraphics[width=\linewidth]{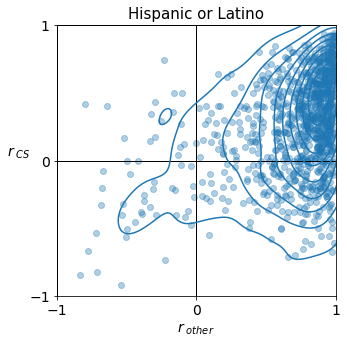}
\end{subfigure}
\vspace{-0.25cm}
\caption{Scatterplot of institutional completion values for CS (vertical axis) compared to other areas of study (horizontal axis). National trends can be further observed using the overlaying, nested regions generated using Kernel Density Estimation (KDE).}
\label{fig:quadrants}
\end{figure*}

\begin{figure}[b!]
\centering
    \includegraphics[width=0.9\columnwidth]{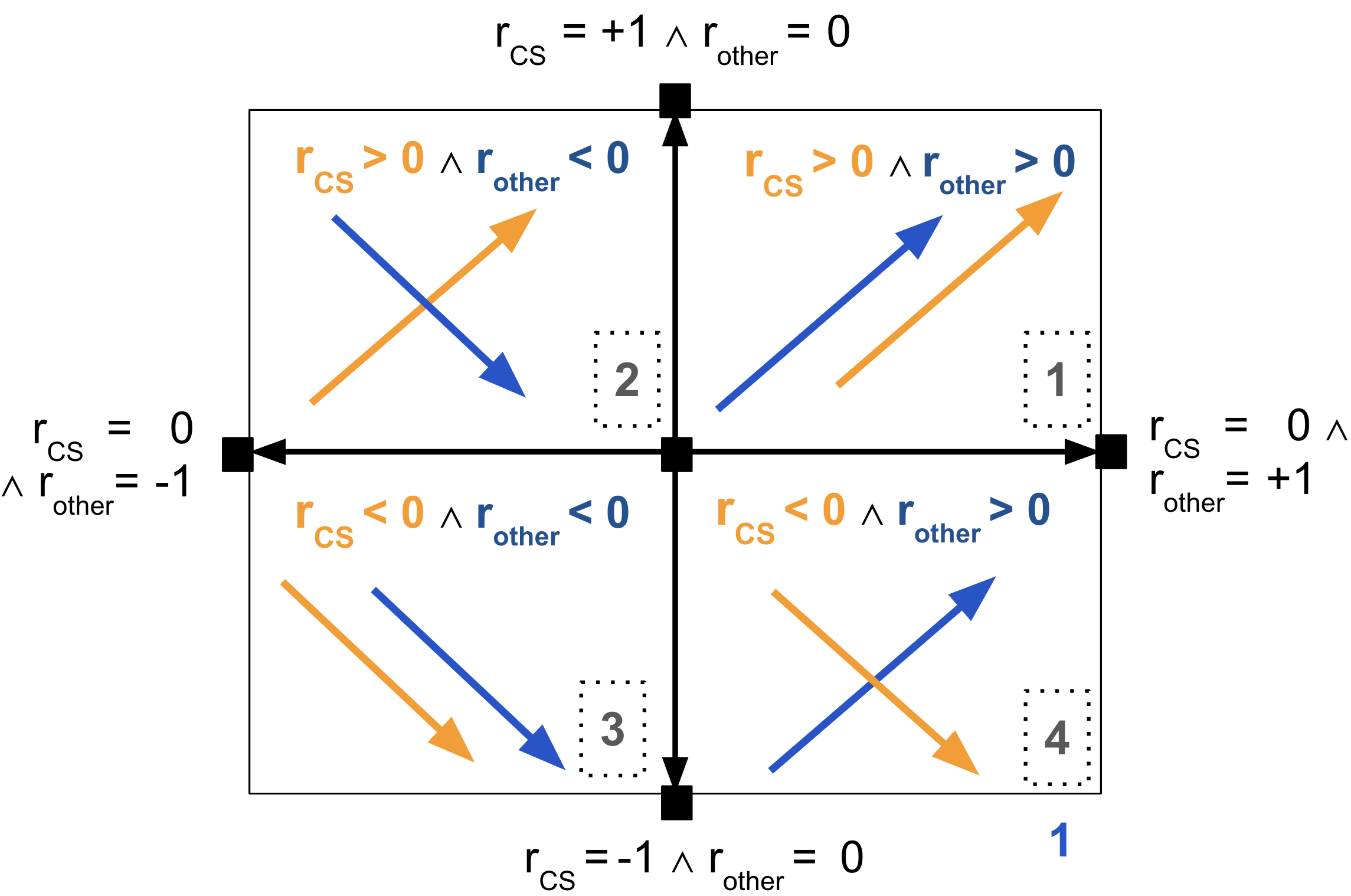}
    \caption{A guide to assist in interpreting the plots in \figref{quadrants}:
    Cartesian plane restricted to $[-1,1] \times [-1,1]$ with $r_{CS}$ as the vertical axis and $r_{other}$ horizontal axis.
    }
    \label{fig:quadrants-legend}
\end{figure}

%
%
\subsection{Lagging Progress}
\label{sec:lagging_progress}

\textbf{National Trends.}
\figref{national_b} presents strong evidence of the large gap between CS and other areas of study when it comes to participation and success of students from underrepresented groups.
For most underrepresented groups, these gaps are only becoming wider with time.
This should be alarming to the CS community.

In 2011, women accounted for only 19\% of the graduating class in CS compared to other areas of study (58\%; Figure 1b). 
While representation of women in the graduating class remained relatively stable at 59\% in non-CS fields in 2019, women accounted for only 21\% of the graduating class in CS.
For black students in CS, the trends are even more bleak.
While CS has much work to do with female student representation, the trend in CS has greater positive strength than other areas of study ($r_{CS \wedge Women} = 0.91 > 0.85 = r_{other \wedge Women}$ summarized in \tableref{diffs}).
For black students, completions in CS \emph{regressed} sharply from 10\% to 8\% of the graduating class.
During the same period, the share of Black completions in other areas of study remained relatively stable at around 9\% of the graduating class.
For reference, 14.2\% of the total US population is Black or African American alone, as per the US Census Bureau~\cite{uscensus}.

For Hispanic or Latino students, trends in both CS and other areas of study are strongly positive, indicating good progress.
However, the positive trend is stronger in other areas of study as compared to in CS $r_{other} = 1.0 > 0.97 = r_{CS}$ (\tableref{diffs}).
This indicates that CS is lagging behind and not able to keep up with other disciplines in its progress for Hispanic or Latino student representation and success.
If these trends (albeit positive) persist, the gap between CS and other areas of study would only widen.

\noindent\textbf{Institutional Trends.}
At the institutional level, we compare completion trends in CS to other areas of study directly for each institution by taking the difference of the two completion trend measures: i.e., $r_{CS} - r_{other}$.
\figref{diffs} depicts the distributions of $r_{CS} - r_{other}$ for each institution and demographic group.
The blue area of each distribution represents institutions where CS completion trends are lagging behind completion trends in other areas of study: $r_{CS} < r_{other}$.
The orange part of each distribution represents institutions where CS trends are better than other areas of study: $r_{CS} > r_{other}$.
\tableref{diffs} summarizes the distributions in \figref{diffs}. 

For Women, we observe in \figref{diffs} that the frequency distribution is a bell-shaped curve with the peak just to the right of zero.
The mean difference (dashed line) for Women is $0.07$ indicating that for an average institution, the completion trends for Women is marginally better in CS than other areas of study.
For Black students, the frequency distribution is similarly bell-shaped, but with a peak to the left of zero.
However, the mean difference (dashed line) for Black students is $-0.06$ indicating that for an average institution, CS is lagging behind other disciplines with respect to completion trends of Black students.

For Hispanic or Latino students, the frequency distribution is also bell-shaped but the peak is far to the left of zero with mean difference of $-0.45$.
This suggests that for an average institution in the United States, CS is lagging far behind other areas of study in terms of completions by Hispanic or Latino students.
We can also observe this result in the rightmost panel of \figref{diffs} in which most of the distribution is blue representing the fact that $r_{CS} < r_{other}$ for an overwhelming majority of institutions.

%
%
\subsection{Simpson's Paradox}
\label{sec:simpson}

In subsections \ref{sec:abs_relative} and \ref{sec:lagging_progress}, we compared completion trends in CS to other areas of study for Women, Black and Hispanic or Latino students.
Our analysis considered data from a nationwide scale (dotted line in Figure~\ref{fig:diffs}) as well as at an institutional scale (dashed line in  Figure~\ref{fig:diffs}).
In some cases, we observe a congruence at both scales and for others, an alarming incongruence.

For Women, the trends at national and institutional scales are congruent: at the national level $r_{CS} - r_{other} = 0.06$ whereas, for an average institution, $\mu_{r_{CS}-r_{other}} = 0.07$, a marginal difference.
However, for Black and Hispanic or Latino students, the trends observed at national level are incongruous with trends observed at institutional level.
We informally observe this incongruity by noting the wide gaps between dotted and dashed lines in the center and right panels of Figure~\ref{fig:diffs}.
Numerically, these gaps are also reflected in divergent national and institutional values in columns 3 and 4 of Table~\ref{table:diffs}.
These data are a clear example of \emph{Simpson's Paradox}: when a trend appears in several groups of data but disappears or reverses when the groups are combined.

It is of particular interest that the effect of this paradox on CS completion trends is different for Black students than for Hispanic or Latino students.
For Black students in CS, the national trend is strongly negative ($r_{CS}-r_{other} = -0.87$) but at the institutional level, the average trend is relatively neutral ($\mu_{~r_{CS}-r_{other}} = -0.06$). 
In contrast, for Hispanic or Latino students, the completion trend observed at the national level is relatively flat ($r_{CS}-r_{other} = -0.03$) whereas the completion trend observed at the institutional level is strongly negative ($\mu_{~r_{CS}-r_{other}} = -0.45$).

%
%

%
%

\noindent\textbf{Unpacking Simpson's Paradox.} To better understand these observed discrepancies between national and institutional trends, we attempt to graphically blend national- and institutional-level data into a single plot for each underrepresented group.
\figref{quadrants} depicts institutional distributions as a scatterplot with respect to completion trends in CS, $r_{CS}$, and completion trends in other areas of study, $r_{other}$.
We simultaneously observe national trends in each of plots in \figref{quadrants} by considering densities identified with the curved regions computed using Kernel Density Estimation (KDE).
To aid the reader, \figref{quadrants-legend} is a legend for how to interpret each plot in \figref{quadrants}.
Our plots are restricted to $[-1,1] \times [-1,1]$ of the Cartesian plane with $r_{CS}$ as the vertical axis and $r_{other}$ horizontal axis crossing at $(0,0)$.
Quadrants are labeled with dotted boxes counterclockwise $1$ through $4$ consistent with mathematical nomenclature.
The colored arrows indicate an upward (orange) or downward (blue) completion trend for a particular institution.
For example, quadrant $2$ indicates $r_{CS} < 0$ and $r_{other}> 0$: the trend of CS completions at an institution is increasing while completions for other areas of study is decreasing.

For all three demographic groups in \figref{quadrants}, the majority of the institutions fall in the quadrant $1$ (top right) where $r_{CS} > 0$ and $r_{other} > 0$.
The KDE distribution for Women is nearly uniformly distributed with peaks in the top-right of quadrant $1$.
This aligns with the mean difference of $0.07$ for Women students observed in \figref{diffs}. 
The region of the plot corresponding to $r_{CS} < -0.5$ is relatively sparse indicating that with respect to CS, very few institutions exhibit strong negative trends.
This is consistent with \figref{national} where completions in other areas of study have stabilized at the saturation point of around $60\%$ but CS still has a lot of ground to cover. 

For Black students (center panel), \figref{quadrants} shows a near bimodal distribution with most institutions in either quadrant $1$ or quadrant $3$, with quadrants $2$ and $4$ being relatively sparse.
This indicates that institutions with a positive trend in CS completions also exhibit a positive trend in other areas of study and vice versa. 
Reconciling the institutional-level scatter-KDE plot with the national level trends we observed in \figref{national_b} suggests that \emph{institutions with $r_{CS} < 0$ for Black students must have a disproportionately large influence to cause a sharply declining trend at the national level}.
We consider this further in subsection \ref{sec:geographic}.

For Hispanic or Latino students, the KDE plot in \figref{quadrants} (right panel) is far more concentrated than the distributions for Women and Black students.
This grouping of institutions in quadrants $1$ and $4$ means the variance along the $r_{CS}$ vertical axis is far greater than the variance along the $r_{other}$ horizontal axis.
Thus, for Hispanic or Latino students, an overwhelming majority of institutions have a strong positive trend for completions in other areas of study while the trends of CS completions varies relatively more from institution to institution.
Reconciling the institutional-level scatter-KDE plot with the national level trends we observed in \figref{diffs} suggests that \emph{institutions with $r_{CS} < 0$ for Hispanic or Latino students must also have a disproportionately large influence to cause a positive trend at the national level but a neutral trend at the institutional level}.

\begin{figure}[t!]
\centering    \includegraphics[width=\columnwidth]{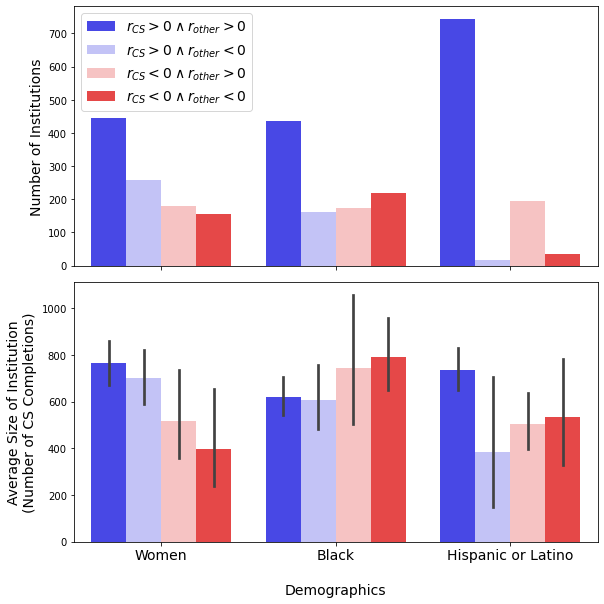}
    \caption{Institutions divided into 4 $r$-categories based on completion trends: values of $r_{CS}$ and $r_{other}$.
    The top panel reflects the number of institutions in each $r$-category for each underrepresented, demographic group.
    The bottom panel depicts the average number of CS completions for institutions  in each $r$-category broken out for each underrepresented, demographic group.
    }
    \label{fig:size}
\end{figure}

%
%
\noindent\textbf{Resolving Simpson's Paradox.}
We attempt to resolve the question of why trends observed at the national level are incongruous with those observed at the institutional level for Black and Hispanic or Latino students. 
\figref{size} divides institutions in our dataset into the 4 $r$-categories corresponding to the quadrants defined in \figref{quadrants-legend}.
The top panel of \figref{size} shows the frequency distribution of institutions and the bottom panel is a frequency distribution with respect to size of institution (where size is based on total CS completions). 

In the top panel of \figref{size} we observe that for every underrepresented demographic group, most institutions fall within quadrant $1$: $r_{CS} > 0$ and $r_{other} > 0$.
It is therefore the case that most institutions exhibit a positive trend in both CS completions and completions in other areas of study.
However, the bottom panel of \figref{size} shows that for Black students, the institutions where $r_{CS} < 0$ (i.e., quadrants $3$ and $4$) also have the largest graduating classes.
This explains why the national-level trends are so different from the institutional-level trends for Black students: \emph{the few institutions that have relatively large CS graduating classes are the ones that exhibit a negative trend in CS completions}.
This contrasts with most institutions that have relatively smaller graduating classes, but exhibit a positive trend in CS completions.
The results are similar for Hispanic or Latino students in \figref{size} where $r_{CS} < 0$ category institutions have a large number of total completions (graduates).
However, because $r_{CS} > 0 \wedge r_{other} > 0$ category institutions have the largest size for Hispanic or Latino students, the institution level trends are neutral or flat as opposed to negative for Black students. 

%
%

\begin{figure*}[t!]
\centering
\captionsetup[subfigure]{justification=Centering}
\begin{subfigure}[t]{.27\textwidth}
    \includegraphics[width=\linewidth]{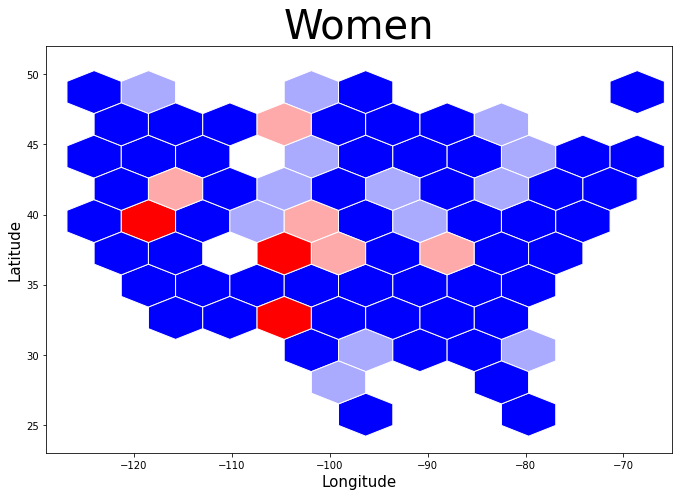}
    \label{map2_w}
\end{subfigure}%
\hspace{\fill} 
\begin{subfigure}[t]{.29\textwidth}
    \includegraphics[width=\linewidth]{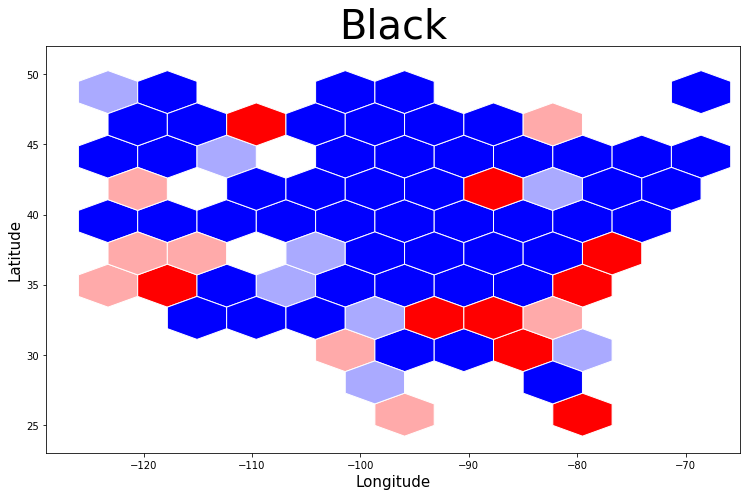}
    \label{map2_b}
\end{subfigure} 
\hspace{\fill} 
\begin{subfigure}[t]{.38\textwidth}
    \includegraphics[width=\linewidth]{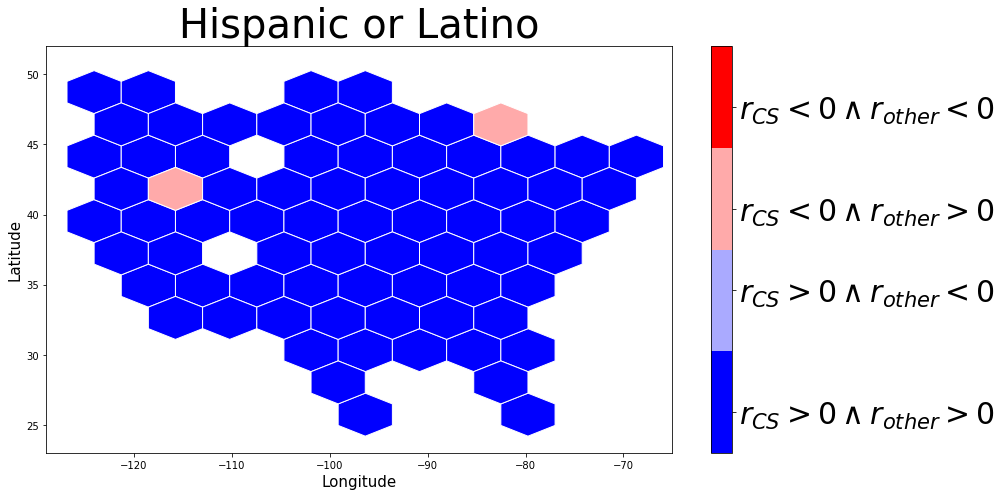}
    \label{map2_h}
\end{subfigure}
\caption{Spatial distribution over the United States of completion trends in CS and other areas of study from 2011 to 2019 for Women, Black, and Hispanic or Latino students.
Each hexagon implies a geographic region and does \emph{not} correspond to a state.}
    \label{fig:maps}   
\end{figure*}

\subsection{Geographic Trends}
\label{sec:geographic}

Thus far, our discussion has focused on completion trends in CS and other areas of study that are strictly temporal.
We augment our analyses with the added dimension of space to identify any spatiotemporal trends in completions for underrepresented students.
\figref{maps} shows hexagon plots over the United States for each demographic group color-coded by completion trends in CS and other areas of study.
White hexagons correspond to areas of the United States with either no institution or institutions with no CS graduates belonging to the given demographic group.
The color of each hexagon is determined by the mode of $r$-category of institutions within the region.
The four categories along with their corresponding color are given in the color bars to the right and are consistent with the $r$-categories colors from \figref{size}.

We observe that completion trends are spatially localized.
For Women, negative trends in CS and other areas of study are observed away from the East and West coasts with a negative cluster near the Central West region.
For Black students, the negative completion trends in CS and other areas of study seem to strongly concentrate in the South.
In particular, we observe a large cluster of negative completion trends in the Southeast.
These negative trends are alarming since the majority of the Black population in the United States lives in these regions, as per the US Census Bureau~\cite{uscensus}.
A strong but relatively smaller negative cluster can also be observed near Southern California and adjacent areas.
For Hispanic or Latino, the US census~\cite{uscensus} shows that the population is concentrated in the Southwest but in contrast to Black students we do not observe any regions of concentrated negative trends.

%
%
\section{Conclusions and Future Work}
\label{seg:conclusion}

We draw the following conclusions from the analyses presented in this work:

\begin{enumerate}
    \item For Women in Computer Science, the completion rates are far behind when compared to other areas of study. However, they are trending in the positive direction.
    
    \item For Black students in Computer Science, we observe a sharp decline in completion trends.
    This is particularly alarming because a similar decline was not observed for black students in other areas of study.
    The negative trend in completions for Black students at the national level occurs despite the fact that most institutions individually exhibit positive trends.
    However, a few institutions with large graduating classes in CS exhibit strong negative trends and are driving most of the national level results.
    Furthermore, the negative trends for Black students are concentrated generally in the South and particularly in the Southeast of the United States.
    
    \item For trends with Hispanic or Latino students, completions in Computer Science and other areas of study are both strongly positive.
    However, the positive trend in CS is still lagging behind other areas of study.
    If this trend continues, the gap between Computer Science and other areas of study will only increase.
    
\end{enumerate}

Taken together, the results presented in this paper serve as a call to action for the CS Education community to include a more data-driven approach to the problem of Diversity, Equity and Inclusion in CS.
There are many potential avenues for future work that expand on the results presented.
For example, while a completion implies that a student is able to overcome the challenges of a computing program, it is still just one measure of success at the end of a long and complex educational pipeline.
Similarly, larger institutions, that seem to have an out-sized influence on national trends in our results, are more likely to be public with low tuition fees and high student-to-faculty ratio. 

In future work, we intend to include more nuanced categories of race and ethnicity and American Indian and Alaska Natives, Native Hawaiian or Other Pacific Islanders and Mixed-race students.
While gender and sexual orientation are important elements of diversity, these are unable to be studied in the same way due to limitations in the way IPEDS data has been collected.
The IPEDS data is nonetheless a rich source of information that can be used to better understand the low participation and success of underrepresented students in CS.
Our work only scratches the surface.

\bibliographystyle{ACM-Reference-Format}
\bibliography{references}

\appendix








\end{document}